\documentclass[jkps,preprint,fleqn,showpacs,showkeys]{revtex4}
\usepackage{graphicx}
\usepackage{amssymb}
\usepackage{amsmath}
\usepackage{bm}
\usepackage{slashed}

\DeclareMathOperator{\tr}{tr}

\newcommand\beq{\begin{eqnarray}}
\newcommand\eeq{\end{eqnarray}}

\begin{document}
\setcounter{page}{0}
\preprint{\parbox[b]{1in}{ \hbox{\tt PNUTP-17/A06}}}
\def\vp{{\vec{p}}}
\def\vx{{\vec{x}}}
\def\vD{{\vec{D}}}
\def\vB{{\vec{B}}}
\def\vE{{\vec{E}}}
\def\vcE{{\vec{\mathcal{E}}}}
\def\vA{{\vec{A}}}
\def\vDel{{\vec{\nabla}}}
\def\vzero{{\vec{0}}}

\def\a{{\alpha}}
\def\b{{\beta}}
\def\d{{\delta}}
\def\D{{\Delta}}
\def\X{{\Xi}}
\def\e{{\varepsilon}}
\def\g{{\gamma}}
\def\G{{\Gamma}}
\def\k{{\kappa}}
\def\l{{\lambda}}
\def\L{{\Lambda}}
\def\m{{\mu}}
\def\n{{\nu}}
\def\o{{\omega}}
\def\O{{\Omega}}
\def\S{{\Sigma}}
\def\s{{\sigma}}
\def\th{{\theta}}

\def\ol#1{{\overline{#1}}}

\def\Dslash{D\hskip-0.65em /}
\def\Dtslash{\tilde{D} \hskip-0.65em /}

\def\CPT{{$\chi$PT}}
\def\QCPT{{Q$\chi$PT}}
\def\PQCPT{{PQ$\chi$PT}}
\def\tr{\text{tr}}
\def\str{\text{str}}
\def\diag{\text{diag}}
\def\order{{\mathcal O}}

\def\cE{{\mathcal E}}
\def\cF{{\mathcal F}}
\def\cG{{\mathcal G}}
\def\cS{{\mathcal S}}
\def\cC{{\mathcal C}}
\def\cB{{\mathcal B}}
\def\cT{{\mathcal T}}
\def\cQ{{\mathcal Q}}
\def\cL{{\mathcal L}}
\def\cO{{\mathcal O}}
\def\cA{{\mathcal A}}
\def\cQ{{\mathcal Q}}
\def\cR{{\mathcal R}}
\def\cs{{\mathfrak s}}
\def\cH{{\mathcal H}}
\def\cW{{\mathcal W}}
\def\cM{{\mathcal M}}
\def\cD{{\mathcal D}}
\def\cN{{\mathcal N}}
\def\cP{{\mathcal P}}
\def\cK{{\mathcal K}}
\def\Qt{{\tilde{Q}}}
\def\Dt{{\tilde{D}}}
\def\St{{\tilde{\Sigma}}}
\def\cBt{{\tilde{\mathcal{B}}}}
\def\cDt{{\tilde{\mathcal{D}}}}
\def\cTt{{\tilde{\mathcal{T}}}}
\def\cMt{{\tilde{\mathcal{M}}}}
\def\At{{\tilde{A}}}
\def\cNt{{\tilde{\mathcal{N}}}}
\def\cOt{{\tilde{\mathcal{O}}}}
\def\cPt{{\tilde{\mathcal{P}}}}
\def\cI{{\mathcal{I}}}
\def\cJ{{\mathcal{J}}}

\def\eqref#1{{(\ref{#1})}}

\title{
Subleading Hadronic Vacuum Polarization Contributions to Muon $g-2$: $\mu^+\mu^- \rightarrow \gamma^* \rightarrow \pi^0 \gamma^*$
}

\author{Deog Ki Hong}
\email[]{dkhong@pusan.ac.kr}
\affiliation{
Department of Physics,
        Pusan National University,  
        Busan 46241, Korea,}
\author{Du Hwan Kim}
\email[]{kdh314@pusan.ac.kr}
\affiliation{
Department of Physics,
        Pusan National University,  
        Busan 46241, Korea,}
\author{Jong-Wan Lee}
\email[]{jwlee823@pusan.ac.kr}
\affiliation{
Department of Physics,
        Pusan National University,  
        Busan 46241, Korea,}
\date{\today}



\begin{abstract}
We consider the subleading contributions of the hadronic vacuum polarization, involving the $\pi^0\gamma^*\gamma^*$ transition form factor,  
to the muon anomalous magnetic moment $g-2$. 
Various models for the form factor, based on hadronic ansatzes and 
holographic principles, are considered:  They are the Wess-Zumino-Witten, vector meson dominance, lowest meson dominance (one and two vector resonances), and anti-de Sitter/quantum chromodynamics (AdS/QCD) models. 
The model parameters are determined by fitting the experimental data for the  $e^+ e^- \rightarrow \pi^0 \gamma$ total cross section. 
We report the following numerical result for the corrections to the muon $g-2$: the resulting values of two vector resonances model 
are one order-of-magntitude smaller than 
the one obtained from the dispersion relation. 
\end{abstract}
\maketitle

\begin{center}
\section{Introduction} %
\label{sec:intro}
\end{center}

The muon anomalous magnetic moment has been playing a crucial role in the precision test of 
the Standard Model (SM) in particle physics, where 
both the theoretical and the experimental values are determined with comparatively high precision 
(see Ref.~\onlinecite{Jegerlehner:2009ry} for a comprehensive review).
There is a considerably large discrepancy between them 
of about $2.4\,\sigma\sim3.6\,\sigma$  deviations, depending on the experimental inputs to the
hadronic vacuum polarization contributions, including the inputs from $e^+e^-$ scattering and $\tau$ decay experiments \cite{Patrignani:2016xqp}. 
The most recent work finds, after accounting for the $\gamma-\rho^0$ mixing to correct the $\tau$ decay analysis, that 
\cite{Jegerlehner:2017lbd} 
\beq
a^{\textrm{exp}}_\mu-a^{\textrm{th}}_\mu=(31.3\pm7.7)\times 10^{-10},
\eeq
which results in a $4.1\sigma$ deviation.
If such a discrepancy actually turns out to persist in the upcoming experiments, it will be strong 
experimental evidence for new physics beyond the SM. 
Therefore, substantial efforts are currently being devoted to 
increasing the precision of both the experimental measurements and the theoretical calculations 
of the muon anomalous magnetic moment. 
For instance, the new experiment at Fermilab plans to reduce the current experimental error by a factor of four \cite{Hertzog:2015jru}.  

\begin{figure}[h]
\begin{center}
\includegraphics[width=0.5\textwidth]{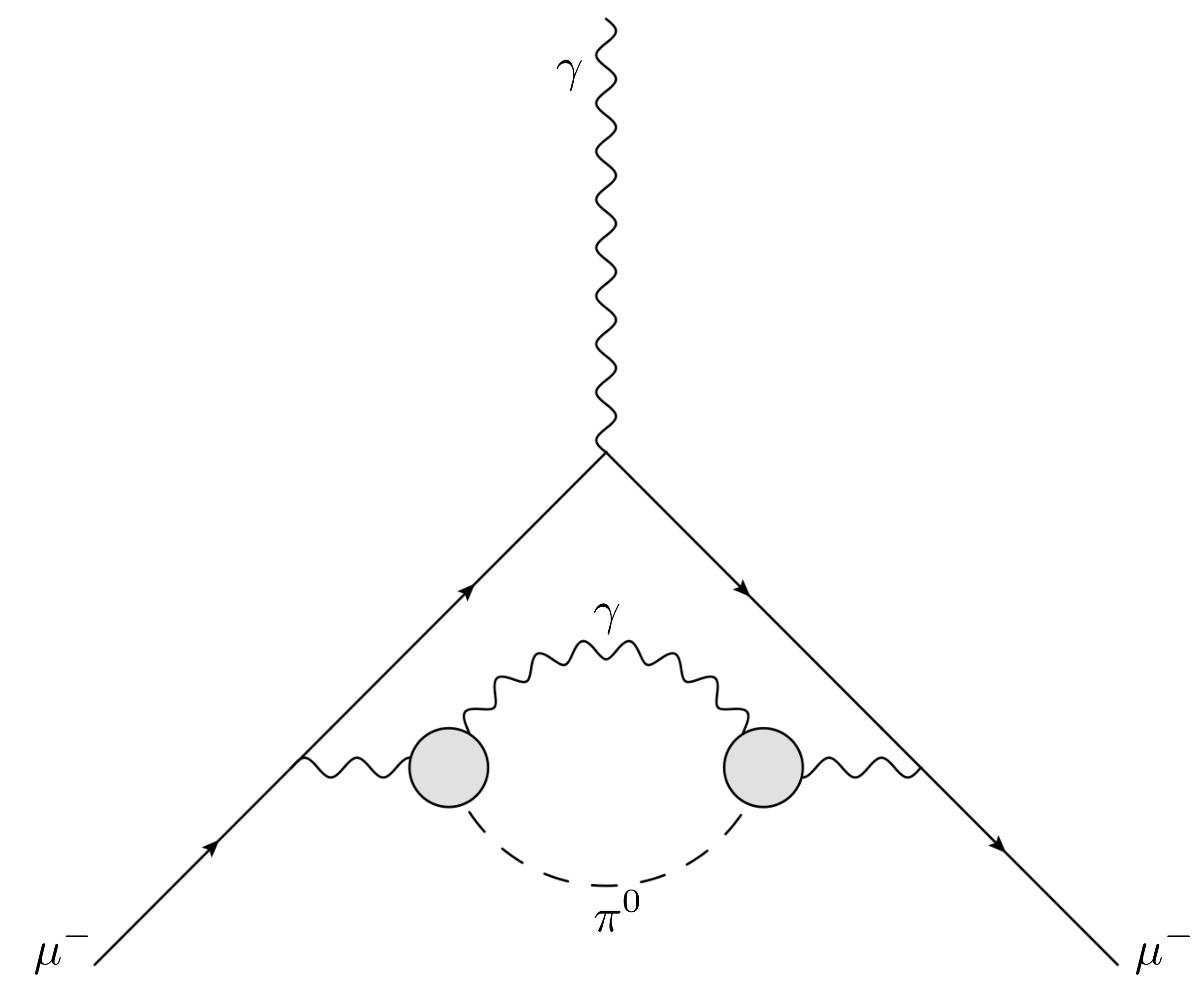}
\end{center}
\caption{Feynman diagram of the HVP contributions to the muon $g-2$ 
involving the anomalous transition, $\gamma^*\rightarrow\pi^0\gamma^*$. 
The grey balls denote the corresponding transition form factors.}
\label{fig:two_loop}
\end{figure}

On the theoretical side, the SM prediction is typically divided into three parts: 
Quantum electrodynamics (QED), electroweak (EW) and hadronic (Had) parts. 
While the QED and EW contributions are determined with unprecedented precision by using perturbation theory \cite{Aoyama:2012wk,Gnendiger:2013pva}, 
the determination of the hadronic part is challenging because of its nonperturbative nature 
and, thus, is responsible for the main theoretical uncertainties. 
Though, in principle, this hadronic part can be computed from first principles by using lattice quantum chromodynamics (QCD), 
it remains currently in an early stage and, thus, involves large statistical 
and systematic errors \cite{DellaMorte:2011aa,Boyle:2011hu,Burger:2013jya,Blum:2014oka,Blum:2015gfa,Chakraborty:2016mwy,Borsanyi:2017zdw}. 
The standard approach to the evaluation of the lowest order hadronic contribution is to use the dispersion relation to obtain the hadronic vacuum polarization (HVP) from the cross-section measurements of the $e^+e^-$ annihilation to hadrons or the $\tau$ decay. Of our particular interest is the subleading HVP for the annihilation channel of $\pi\gamma^*$, which is at the order of $\alpha^3$.
The corresponding Feynman diagram is shown in Figure \ref{fig:two_loop}. 
In fact, this two-loop diagram contains the double-virtual pion-photon-photon transition form factor, 
$F_{\pi^0\gamma^*\gamma^*}(q_1^2,q_2^2)$, which is used to calculate the hadronic light-by-light (HLBL) 
contribution in the pion pole approximation \cite{Knecht:2002,Nyffeler:2016gnb}. 
This transition form factor largely depends on the hadronic models 
based on the vector meson dominance and is constrained by the large-$N_c$ QCD and the operator product expansion. 
Recently, an alternative technique for the determination of the form factor is introduced by using 
holographic principles~\cite{Hong:2009,Hong:2009jv,Cappiello:2010uy}. 
The purpose of this paper is to calculate the HVP contributions to the muon $g-2$ that involves the anomalous transition  
$\gamma^*\rightarrow \pi^0\gamma^*$.
We compare our results with the known values obtained from the dispersive approaches \cite{Davier:2017zfy}
and discuss their implications for the determination of the HLBL contributions. 

This paper is organized as follows: 
In Section \ref{sec:HVP}, we define the HVP matrix element for the $\pi^0\gamma^*$ channel,  
where the pion-photon-photon transition form factor is defined in the Lagrangian for quarks and photons. 
Section \ref{sec:form_factor} is devoted to describing the transition form factors 
in various phenomenological models. 
In Section \ref{sec:result}, we first determine the model parameters 
by fitting recent experimental data for the total cross section of $e^+e^-\rightarrow \pi^0\gamma$ 
to those obtained by the models. 
Then, we report our final results for the subleading HVP contributions to the anomalous magnetic moment of muon 
and discuss them critically. 

\section{Evaluation of the hadronic vacuum polarization for $\gamma^*\rightarrow\pi^0\gamma^*$} 
\label{sec:HVP} %

The generic hadronic vacuum polarization contribution to the anomalous moment of the muon, $a_\mu^{\rm HVP}$, 
can be obtained by inserting the vacuum polarization into the leading order QED diagram. 
Then, it is straightforward to show that \cite{Blum:2002}
\beq
a_{\mu}^{\rm HVP}={\Bigl (}\frac{\alpha}{\pi}{\Bigl
  )}^2\int^{\infty}_{0}dK^2f(K^2){\bar \Pi}(K^2),
\label{eq:a_HVP}
\eeq
where $K$ is the Euclidean momentum and $\alpha$ is the fine structure constant. 
The kernel $f(K^2)$ is given as
\beq
f(K^2)=\frac{m_\mu^2 K^2 Z^3 (1-K^2 Z)}{1+m_\mu^2 K^2 Z^2},
\eeq
where $m_\mu$ is the mass of the muon and $Z=-\left[K^2-(K^4+4m_\mu^2 K^2)^{1/2}\right]/2m_\mu^2 K^2$. 
The hadronic vacuum polarization in QCD is defiend by the electromagnetic currents of the quarks as 
\beq
i\int d^4 x e^{iqx} \langle 0 | T \{J^\mu_{\textrm{em}}(x) J^\nu_{\textrm{em}}(0)\}|0\rangle=(q^2\eta^{\mu\nu}-q^\mu q^\nu)\Pi(-q^2),
\eeq
where $J^\mu_{\mathrm{em}}=\sum_f \bar{q}_f\gamma^\mu Q_{\textrm{em}} q_f$ with the quark flavors $(f=u,d,s,\cdots)$ 
and $Q_{\mathrm{em}}$ is the electric charge operator. 
Our convention for the metric tensor is $\eta^{\mu\nu}=\textrm{diag}(1,-1,-1,-1)$. 
The vacuum polarization should be properly regularized and renormalized to handle the ultra-violet (UV) divergences 
in the loop integrals. In this work, we use the hard momentum cut-off $\Lambda_{\mathrm{cut}}$ as a regulator. 
To take account of the charge renormalization, one has to subtract the vacuum polarization at zero momentum 
from the bare vacuum polarization. 
In Eq. \ref{eq:a_HVP}, we, therefore, use the renormalized vacuum polarization function $\bar{\Pi}(K^2)$ given by
\beq
\bar{\Pi}(K^2)=4\pi^2\left[\Pi(K^2)-\Pi(0)\right].
\eeq

\begin{figure}[h]
\centering
\includegraphics[width=0.79\textwidth]{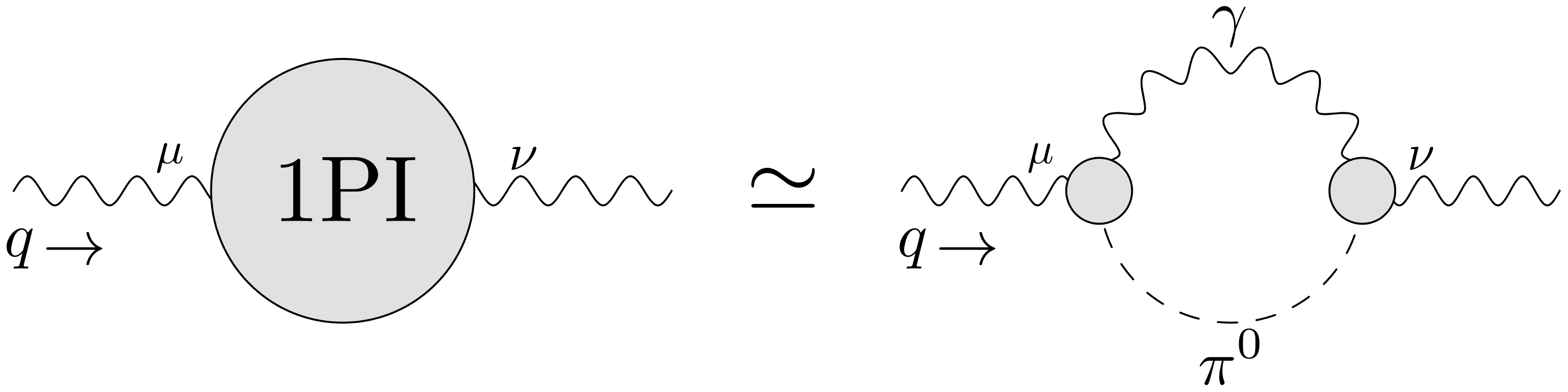}
\caption{
The one-particle irreducible Feynman diagram for the vacuum polarization $\Pi_{\pi^0\gamma^*}$.
}
\label{fig:1PI}
\end{figure}

Because we are considering the hadronic vacuum polarization due to the anomalous transition form-factor of pions, shown in  Figure \ref{fig:1PI}, 
we should rewrite $\Pi(K^2)$ in terms of the anomalous form-factor, $F_{\pi^0 \gamma^* \gamma^*} (q_1^2,q_2^2)$. 
This form-factor basically describes the interaction of the pion with two (off-shell) photons with momenta $q_1$ and $q_2$, respectively.
The interaction Lagrangian for the photons, quarks and muon fields in the underlying theory is given as
\beq
{\cal L}_{\rm int}=-e {\bar {u}}_{\mu}{\slashed A} u_{\mu}+\sum
  _fQ_{em}{\bar q}_{f}\gamma^{\nu}q_{f}A_{\nu},
\eeq
where $u_{\mu}$ denotes the muon field. Starting from this Lagrangian and the QCD Lagrangian,
one can express the anomalous form factor $F_{\pi^0 \gamma^* \gamma^*} (q_1^2,q_2^2)$ as, 
by sandwiching two electromagnetic currents of quarks between the QCD vacuum and the pion state,
\beq
\int d^4x e^{i q_2 \cdot x}\left<0\right|T\{j_{\sigma,\mathrm{em}}(x)j_{\rho,\mathrm{em}}(0)\}\left|\pi^0(q_2-q_1)\right>
=i F_{\pi^{0}\gamma*\gamma*}(q_1^2,q_2^2) \epsilon_{\mu\nu\sigma\rho} q_1^{\mu}q_2^{\nu},
\eeq
where $F_{\pi^{0}\gamma*\gamma*}(q_1^2,q_2^2)=F_{\pi^{0}\gamma*\gamma*}(q_2^2,q_1^2)$ 
as the two photons are indistinguishable.

The one-particle irreducible ($1\textrm{PI}$) diagram of the vacuum polarization for $\gamma^*\rightarrow\pi^0\gamma^*$ 
is given in Figure \ref{fig:1PI}, where the momentum $q$ and the Lorentz indices are understood. 
Then, the $1\textrm{PI}$ matrix element can be expressed as
\beq
i{\cal{M}}^{\alpha\delta}_{\pi^0\gamma}(q^2)=i\int_l
F_{\pi^0\gamma*\gamma*}(q^2,q_-^2)
F_{\pi^0\gamma*\gamma*}(q_-^2,q^2)
\frac{i\epsilon^{\mu'\nu'\delta\gamma}(q_-)_{\mu'}q_{\nu'}(-i g_{\beta\gamma})
i\epsilon^{\mu\nu\beta\alpha}(q_-)_{\mu}q_{\nu}}{[q_-^2+i\epsilon][\ell^2-m^2_{\pi^0}+i\epsilon]}\,,
\label{eq:1PI}
\eeq
where $q_-=q-l$. 
To evaluate the momentum integral, we use the Feynman parametrization and perform the Wick rotation 
to the Euclidean space. For the various models used to study the pion transition form factor, 
we first check that the matrix elements satisfy the Ward-Takahashi identity 
to give the vacuum polarization functions as
\beq
i{\cal M}^{\alpha\delta}_{\pi^0\gamma}
=(q^2\eta^{\alpha\delta}-q^{\alpha}q^{\delta})\Pi_{\pi^0\gamma}(q^2).
\eeq

\begin{center}
\section{anomalous pion-photon-photon form factor} 
\label{sec:form_factor} 
\end{center}%

In the literature, various models for the anomalous pion-photon-photon form factor exist. 
The simplest model might be the model involving Wess-Zumino-Witten (WZW) term \cite{deRafael:1993za}, 
which describes the constraints from the Adler-Bell-Jackiw (ABJ) anomaly at low energy \cite{Adler:1969gk,Bell:1969ts}, 
where the pion decay constant in the chiral limit is replaced by the physical one $F_\pi=92.4~ \textrm{MeV}$. In this model, the form factor is given by 
\beq
F^{\textrm{WZW}}_{\pi^0\gamma^*\gamma^*}(q_1^2,q_2^2)
=F^{\textrm{WZW}}_{\pi^0\gamma^*\gamma^*}(0,0)
=\frac{N_c}{12\pi^2F_{\pi}},
\label{eq:WZW}
\eeq
where $N_c$ is the number of colors in QCD, i.e., $N_c=3$. 
Although this model is not realistic because it does not depend on the momentum, 
it will serve as  a reference point for the theoretical normalization of the anomalous form-factor at $q_1=q_2=0$. 

The next model considered in this work is the usual vector meson dominance (VMD) model, where the form factor is 
given as  
\beq
F^{\textrm{VMD}}_{\pi^0\gamma^*\gamma^*} (q_1^2,q_2^2)=\frac{N_c}{12\pi^2F_{\pi}}\frac{M_V^2}{(q_1^2-M_V^2)}\frac{M_V^2}{(q_2^2-M_V^2)},
\label{eq:VMD}
\eeq
where $M_V$ is the mass of the lightest vector meson and is typically set to be the $\rho$ meson mass. 
The zero momentum limit of this form factor satisfies the axial anomaly constraints. 
Another constraint from QCD is that all the anomalous form-factors should satisfy the requirement that at large Euclidean (spacelike) momenta, the single- and the double-virtual photon form factors 
can be computed using the operator product expansion (OPE), which leads to the so-called Brodsky-Lepage formula 
or equivalently $F_{\pi^0\gamma^*\gamma^*}(-K^2,0)\sim F_{\pi^0\gamma^*\gamma^*}(-K^2,-K^2)\sim 1/K^2$ as $K^2\rightarrow \infty$ 
\cite{Lepage:1979zb,Lepage:1980fj,Brodsky:1981rp,Novikov:1983jt}. 
The VMD model shows a behavior consistent with the Brodsky-Lepage formula for the single-virtual form factor but not for the double-virtual one.
 
We consider two more phenomenological models based on the large-$N_c$ approximation to QCD: 
the lowest meson dominance (LMD) model involving one vector resonance \cite{Moussallam:1994xp,Knecht:1999gb} 
and its variant model (LMD+V) involving two vector resonances \cite{Knecht:2001xc}. 
The form factors are parametrized as
\beq
&{}&F^{LMD}_{\pi^0\gamma^*\gamma^*}(q_1^2,q_2^2)
=-\frac{F_\pi}{3}\frac{q_1^2+q_2^2-c_V}{(q_1^2-M_V^2)(q_2^2-M_V^2)}, \\
&{}&F^{LMD+V}_{\pi^0\gamma^*\gamma^*} (q_1^2,q_2^2)
=-\frac{F_\pi}{3}\frac{q_1^2q_2^2(q_1^2+q_2^2)+h_1(q_1^2+q_2^2)^2+h_2q_1^2q_2^2+h_5(q_1^2+q_2^2)+h_7}
{(q_1^2-M_{V_1}^2)(q_1^2-M_{V_2}^2) (q_2^2-M_{V_1}^2)  (q_2^2-M_{V_2}^2)}, 
\label{eq:LMD}
\eeq
where the overall sign is opposite to that of the anomalous form-factor used in Ref.~\onlinecite{Knecht:2002}. 
The normalization at zero momenta in Eq. \ref{eq:WZW} allows us to 
find the expressions for $c_V$ and $h_7$: 
\beq
c_V=\frac{N_c M_V^4}{4 \pi^2 F_\pi^2} ~~\textrm{and}~~
h_7=-\frac{N_c M_{V_1}^4 M_{V_2}^4}{4 \pi^2 F_\pi^2}.
\eeq
Furthermore, the prefactor of $-F_\pi/3$ is set to be consistent with the leading-order OPE prediction 
for double-virtual photons at short distances. 
Because the single-virtual form factor in the LMD model converges to a constant at large Euclidean momentum, 
it fails to reproduce the Brodsky-Lepage behavior. 
The LMD+V model provides the most promising form factor so far, as it does satisfy 
all the constraints mentioned above if the parameters, the $h_i$'s, are determined appropriately. 
First of all, the large momentum limit of the single-virtual form factor is consistent with 
the Brodsky-Lepage behavior if $h_1$ is set equal to zero and the value of $h_5$ is determined by experiment. 
For instance, the parameter $h_5$ has been set by fitting the single-virtual form vector to the CLEO data \cite{Knecht:2001xc}, which yields
$h_5=(6.93\pm0.26)\,\textrm{GeV}^4$. 
In addition, one determines the parameter $h_2\simeq-10.63\,\textrm{GeV}^2$ by investigating the subleading-order terms (higher-twist corrections) 
in the OPE of the double-virtual form factor at short distances \cite{Melnikov:2003xd}.
A similar result has been obtained from a recent lattice QCD calculation for the form factor to find $h_2=-11.2(5.4)(2.7)\,\textrm{GeV}^2$, 
where the numerics in the first and the second parentheses are the statistical and the systematic errors, respectively \cite{Gerardin:2016cqj}. 
In the LMD+V model, $\rho'$ ($1450$) is often considered for the second lightest vector resonance, if restricted to two light quark-flavors.

\begin{figure}[h]
\centering
\includegraphics[width=0.49\textwidth]{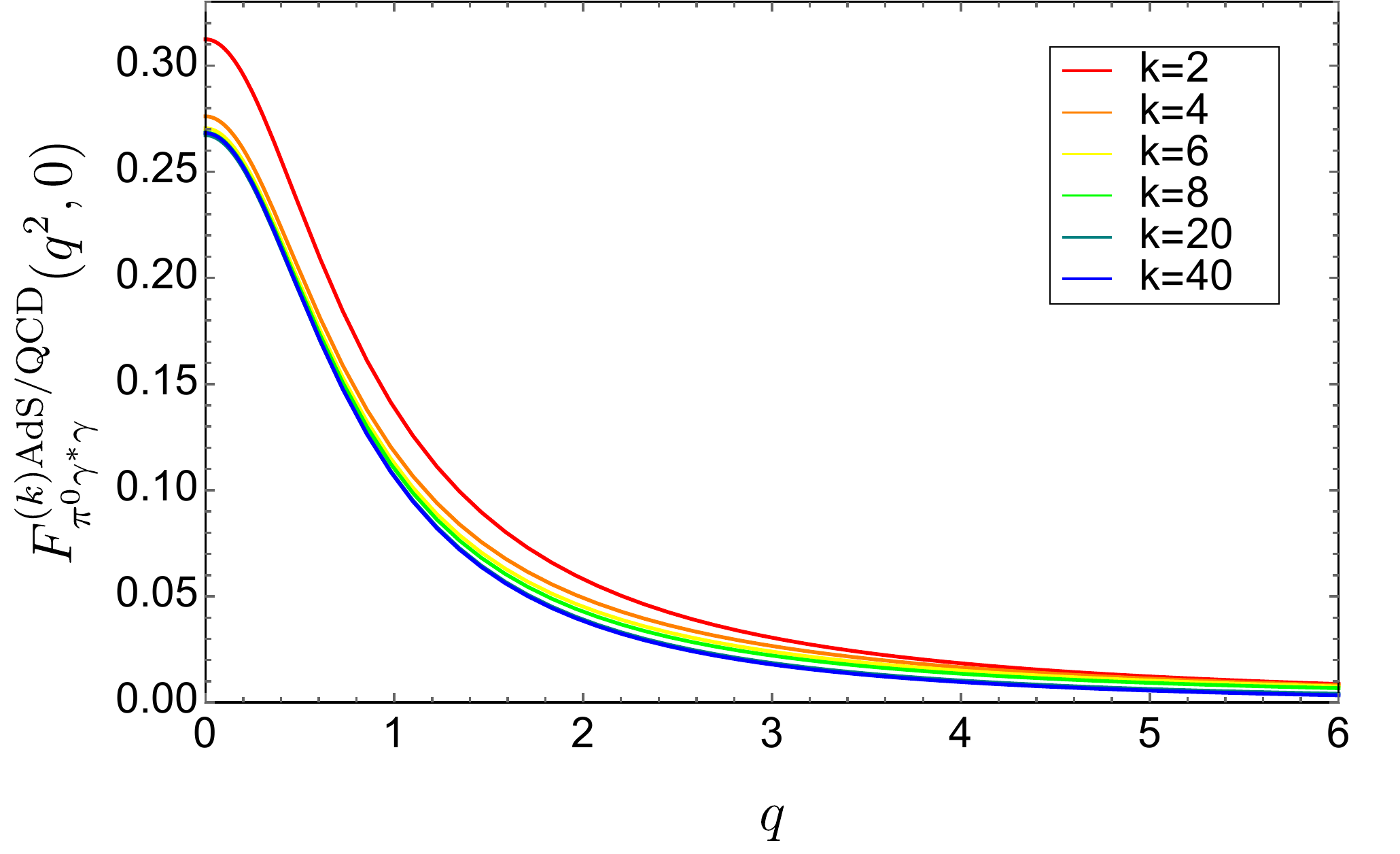}
\caption{Truncated form-factor 
$F^{(k)\rm AdS/QCD}_{\pi^0\gamma*\gamma}(q^2,0)$ for various values of $k$.}
\label{amf}
\end{figure}

\begin{figure}[h]
\centering
\includegraphics[width=0.49\textwidth]{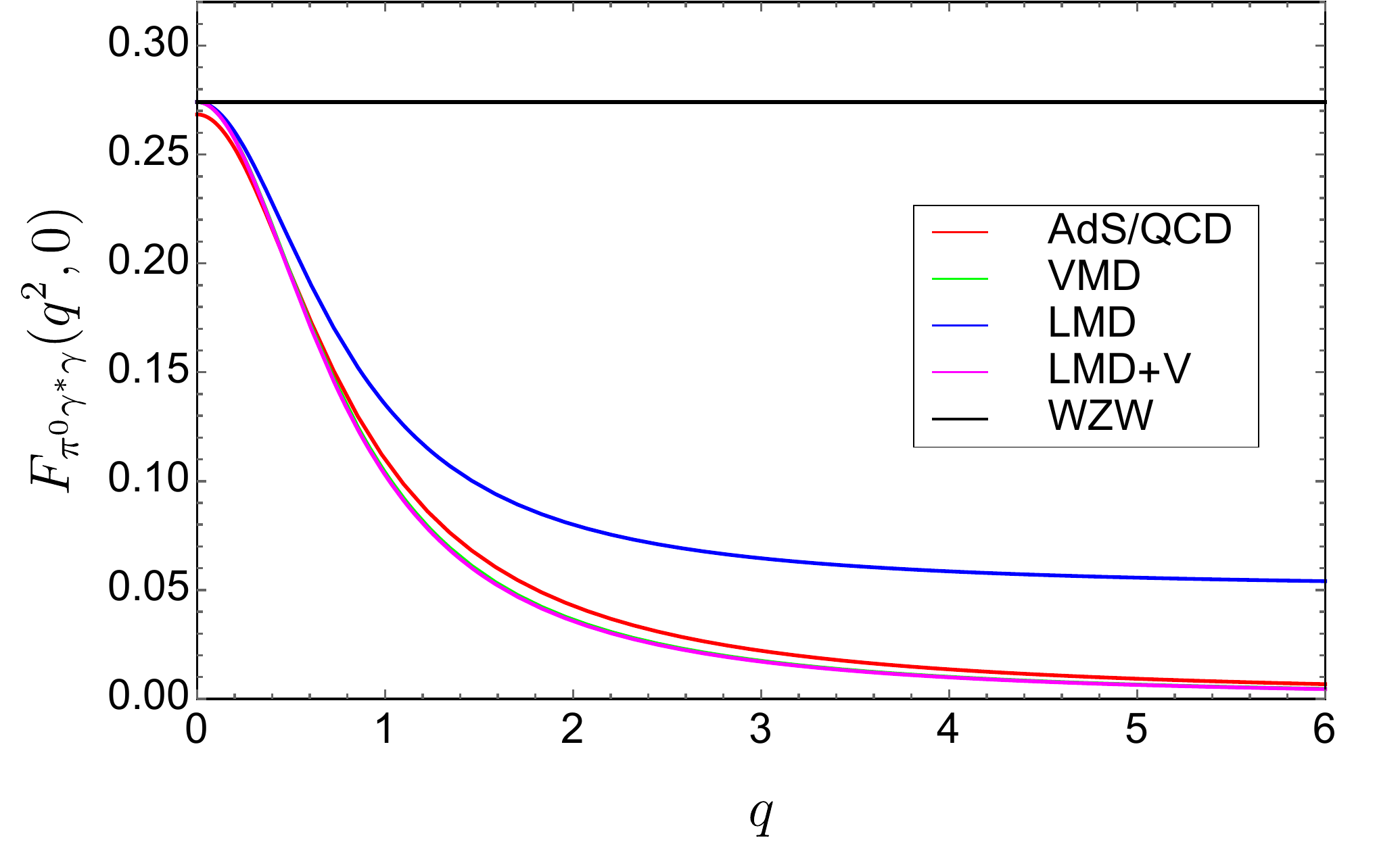}
\includegraphics[width=0.49\textwidth]{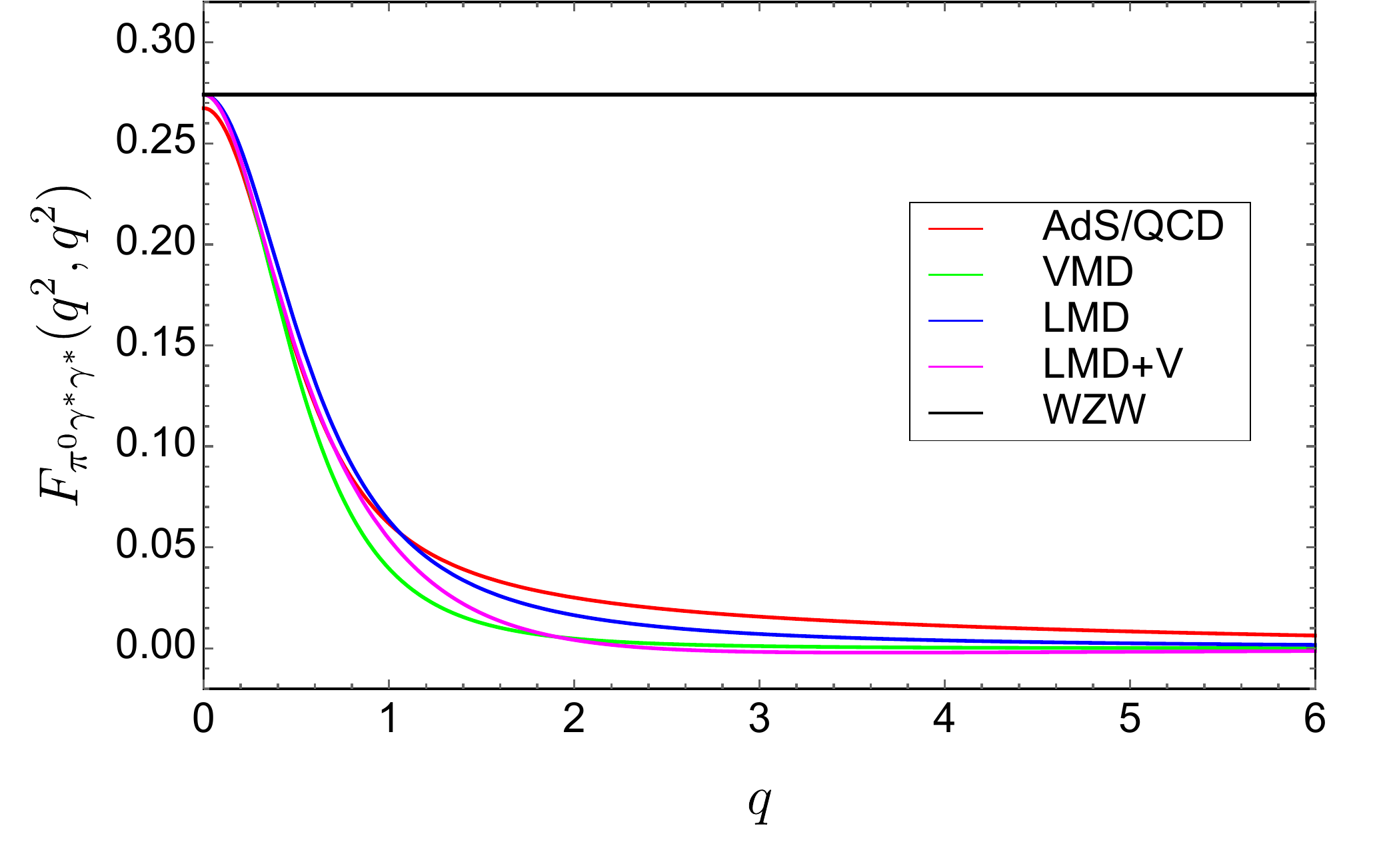}
\caption{The left panel shows the shape of
$F_{\pi^0\gamma*\gamma}(q^2,0)=F_{\pi^0\gamma\gamma*}(0,q^2)$. At infinity, the LMD (blue line) converges to a larger value, compared to the other models.  In the right panel, we plot the anomalous form-factor at symmetric momenta, 
$F_{\pi^0\gamma*\gamma*}(q^2,q^2)$, for each model. 
 }
\label{lmd}
\end{figure}

An alternative approach to determine the transition form factor was recently introduced 
in the framework of anti-de Sitter/QCD (AdS/QCD) as \cite{Hong:2009, Grigoryan:2008, Hovhannes:2007} 
\begin{eqnarray}
F_{\pi^0\gamma*\gamma*}^{\rm AdS/QCD}(q_1^2,q_2^2)=\frac{N_C}{12\pi^2}[\Psi(z_m)J(q_1,z_m)J(q_2,z_m)-\int^{z_m}_{0} dz \partial_z\Psi(z)J(q_1,z)J(q_2,z)]\,,
\label{amf_full}
\end{eqnarray}
where $z_m$ is the infrared cut-off in AdS/QCD that describes the confinement in QCD. $J(q,z)$ and $\Psi(z)$, which describe the non-normalizable external photon source and the pion wave-function, respectively,   are
composed of Bessel functions and modified Bessel functions. Because
$J(q,z)$ is hard to handle, we replace it with a series expansion by using the completeness relation of the (vector meson) wave-functions as introduced in Ref.~\onlinecite{Hong:2004}:
\begin{eqnarray}
J(q,z)=g_5\sum^{\infty}_{k=1}\frac{f_{k}\psi_{k}(z)}{-q^2+M^2_{k}},
\end{eqnarray}
with
\begin{eqnarray}
\psi_{k}(z)=\frac{\sqrt 2}{z_{m}J_{1}(\gamma_{0,k})}zJ_{1}(M_{k}z),
\end{eqnarray} 
which is to satisfy the boundary conditions
$\psi_{k}(0)=0=\partial_{z}\psi(z_m)$. $\gamma_{0,k}$ are the zeros
of the Bessel function $J_{0}(x)$, and the vector meson mass
$M_{k}$ obeys
$M_{k}=\gamma_{0,k}/z_{m}$. The $k$-th meson decay constants, $f_{k}$, are 
determined as
\begin{eqnarray}
f_k=\frac{1}{g_5} {\Bigl [}\frac{1}{z}\partial_z\psi_k(z) {\Bigl ]}_{z=0}=\frac{{\sqrt 2}M_k}{g_5z_mJ_1(\gamma_{0,k})}.
\end{eqnarray}
Using the above results, we truncate the anomalous form factor,  Eq. (\ref{amf_full}), up to the $k$-th vector mesons to obtain 
\begin{eqnarray}
F^{(k)\rm AdS/QCD}_{\pi^0\gamma^*\gamma^*}(q_1^2,q_2^2)=\sum _{i,j=1}^k\frac{c_{ij}}{(q_1^2-M_i^2)(q_2^2-M_j^2)},
\label{eq:adsqcd_formfactor}
\end{eqnarray}
with
\begin{eqnarray}
\!\!\!c_{k\ell}=\frac{N_C}{12\pi^2} {\Bigl [} \Psi(z_m) (g_5 f_k \psi_k(z_m)) (g_5 f_{\ell} \psi_{\ell}(z_m))-\int^{z_m}_{0}\!\!\!{\rm d}z\,\partial_z\Psi(z) (g_5 f_k \psi_k(z)) (g_5 f_{\ell} \psi_{\ell}(z))  {\Bigl ]}.
\end{eqnarray}
In our calculation, we sum the modes up to the 8-th vector meson in Eq.~\ref{eq:adsqcd_formfactor}, because the sum converges rather quickly and the higher excited mesons contribute little (see Figure~\ref{amf}.) 
Depending on the model, the form factors vary slightly, but have
similar shapes, as shown in Figure~\ref{lmd}. Using the form
factor that we obtained above, we calculate the vacuum polarization to find the next-leading hadronic vacuum polarization contribution to 
the muon anomalous magnetic moment $g-2$.\\
\\

\begin{center}
\section{Results and discussion}
\label{sec:result}
\end{center}

If the subleading HVP corrections to the muon $g-2$ are to be calculated
by modeling the pion transition form factors, 
experimental inputs are needed to fix the model parameters. 
In principle, these parameters can also be determined from first-principles calculations 
such as there is lattice QCD. 
In this work, we attempt to determine some of the phenomenological parameters 
by considering recent experimental results for the total cross section of $e^+e^-\rightarrow\gamma^*\rightarrow\pi^0\gamma$ 
from the spherical neutral detector (SND) experiments \cite{SND:2016}. 
For a given single-virtual form factor, $F_{\pi^0\gamma^*\gamma}(q^2,0)$, the total cross section is  given as 
\beq
\sigma_{\textrm{total}}^{e^+ e^- \rightarrow \pi^0 \gamma}=
\frac{2\pi^{2}\alpha^{3}|F_{\pi^0\gamma^*\gamma}(s,0)|^{2}(s+2m_{e}^{2})(s-m_{\pi}^{2})^{3}}{3s^{3}{\sqrt {(s^{2}-4m_{e}^{2}s)}}},
\label{eq:total_cross}
\eeq
where $m_e$ is the electron mass and $s$ is the center-of-mass energy squared. 
From the SND data in the range $0.6\, \textrm{GeV}\leq \sqrt{s} \leq 1.06\, \textrm{GeV}$, 
one can identify two peaks, one at $782.4\,\textrm{MeV}$ and the other at $1018.7\,\textrm{MeV}$, 
corresponding to $\rho$ and $\phi$ mesons, respectively. 
The parameters of the WZW, VMD, and LMD models are completely fixed if the mass of the lightest vector meson, i.e., $\rho$, is given. 
On the other hand, $h_5$ in the LMD+V model can be determined by fitting the data to the total cross section in Eq. \ref{eq:total_cross} 
with $F_{\pi^0\gamma^*\gamma}^{\textrm{LMD+V}}$. 
Note that $h_1$ and $h_7$  in the LMD+V model are fixed by the axial anomaly constraints and the Brodsky-Lepage formula 
while $h_2$ cannot be determined from the SND data, because one of the photons is on-shell. 

\begin{figure}[h]
\centering
\includegraphics[width=0.49\textwidth]{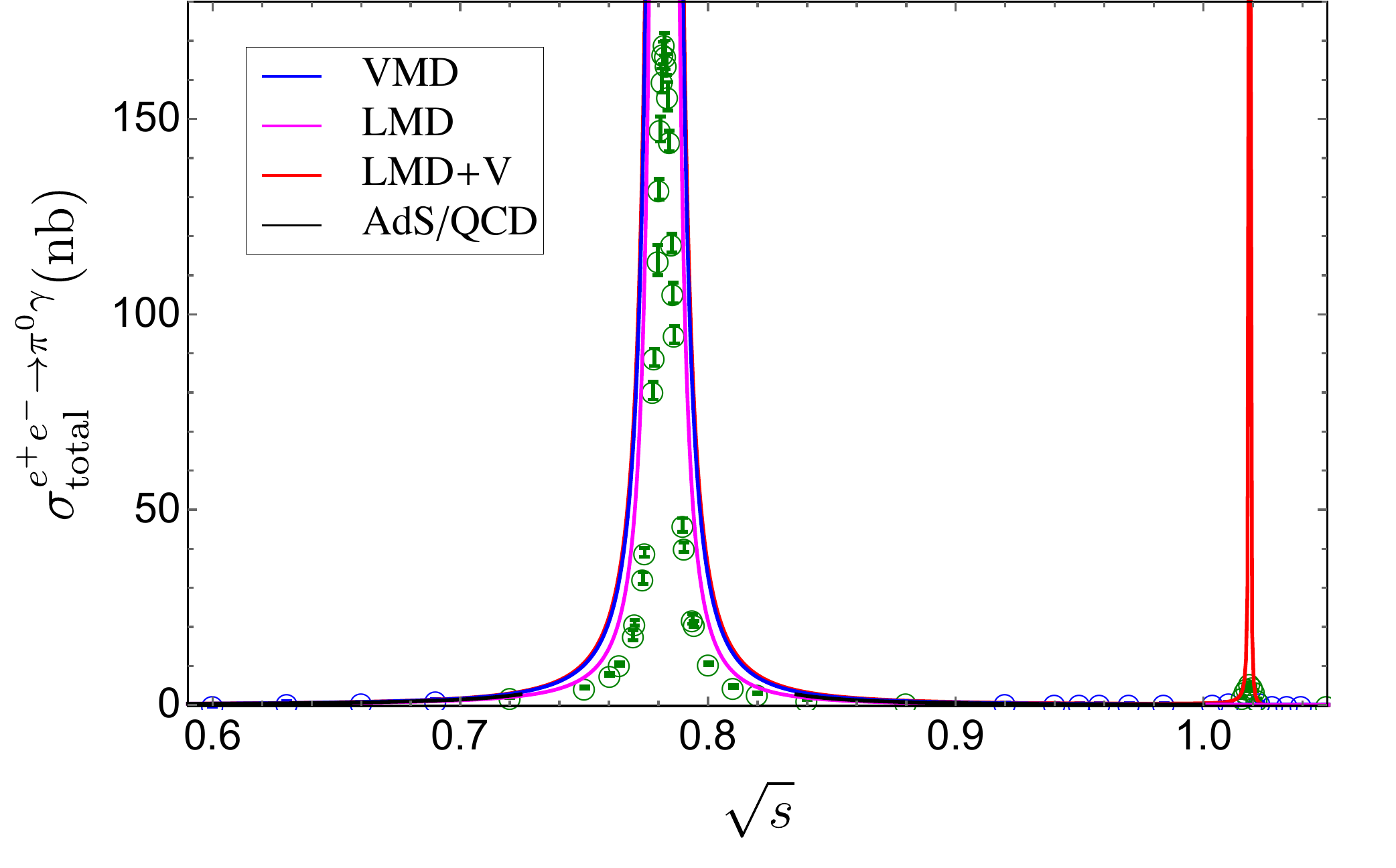}
\includegraphics[width=0.49\textwidth]{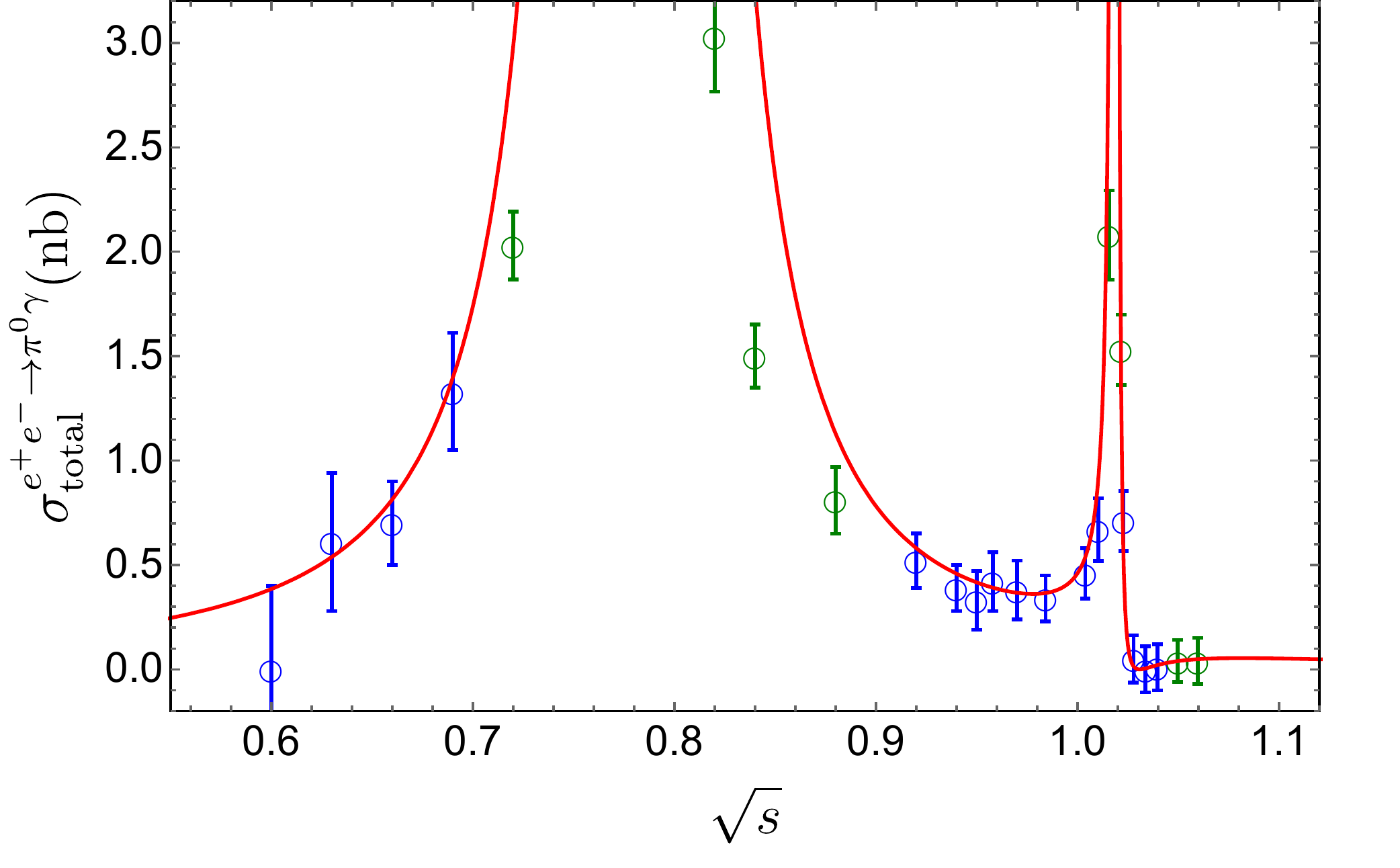}
\caption{
(left) Total cross section defined in Eq.~\ref{eq:total_cross} with
the form factors for the VMD, LDM, LMD+V and AdS/QCD models 
along with the SND data. 
(right) The SND data for the total cross section with the fitted result of the LMD+V model. 
For the fits, we use only the blue data points. 
}
\label{fig:total_cross}
\end{figure}

Because the phenomenological models for the form factor use only the information at the mass pole of the vector meson 
without its decay width, the cross section around the pole cannot be described correctly. 
The cross section diverges at the pole, as shown in the left panel of Figure \ref{fig:total_cross}. For the models in Figure~\ref{fig:total_cross}, we set the mass of the lightest vector meson to $M_V=m_\rho=782.4\,\textrm{MeV}$. 
In the case of the LMD+V model, we perform a single parameter fit to the data far from the poles (blue data points), 
over the ranges $0.6\,\textrm{GeV}\leq\sqrt{s}\leq 0.7\,\textrm{GeV}$, 
$0.9\,\textrm{GeV}\leq\sqrt{s}\leq 1.01\,\textrm{GeV}$ and $1.02\,\textrm{GeV}\leq\sqrt{s}\leq 1.04\,\textrm{GeV}$.
As a result, we obtain $h_5=3.38\,\textrm{GeV}^4$ with $\chi^2/\textrm{d.o.f}\sim 0.5$. 
In the right panel of Figure \ref{fig:total_cross}, we show the fitted results to the SND data for the total cross section. 
We set the parameter $z_m$ appearing in the form factor of the AdS/QCD model to 
$z_m=1/0.3253$ to match the mass of the lightest vector meson to the position of the first peak in the SND data. 

Using the above parameters for the form factors in Eqs. \ref{eq:WZW}-\ref{eq:LMD} and \ref{eq:adsqcd_formfactor} 
and Eq. \ref{eq:a_HVP}, we calculate the  HVP corrections from the vacuum polarization $\pi^0\gamma^*$ to the muon $g-2$. 
We first consider the WZW, VMD, LMD and AdS/QCD models, for which the results are summarized in Table \ref{tab:g2_table}. 
For these models only the lowest vector meson mass is required to fix the model parameters. 
We vary the cut-off scale from $1\,\textrm{GeV}$ to $3\,\textrm{GeV}$. Note that the loop integrals for the VMD and the AdS/QCD models do not have an UV divergence. 
Therefore, the corresponding results for the muon $g-2$ contributions converge to a constant as $\Lambda_{\textrm{cut-off}}$ approaches infinity. 
The HVP corrections for the other two models largely depend on the cut-off scales. 
Although there are considerable model and cut-off dependences, the overall sizes of the corrections have the same orders of magnitude, $\sim 10^{-11}$. 

For the LMD+V model, we use two additional parameters, $h_5$ and
$M_{V_2}$, determined by fitting the SND data, as described above. 
As a reference, we also calculate the HVP contributions to the muon $g-2$ by using the value of $h_5$ used in the HLBL calculation 
and by taking the same values of $M_{V_1}$ and $M_{V_2}$ corresponding
to the two peaks in the SND data. 
Furthermore, we vary the cut-off scale from $1\,\textrm{GeV}$ to
$3\,\textrm{GeV}$ and the parameter $h_2$ by $\pm 10\,\textrm{GeV}^2$. 
Note that $h_2$ cannot be determined if any one of the photons of the pion-photon-photon transition is on-shell. 
Thus, we borrow the value of $h_2=-10.63\,\textrm{GeV}^2$ from Ref.~\onlinecite{Knecht:2002} and consider two more values of $0$ and $10\,\textrm{GeV}^2$ to estimate its dependence on the HVP contributions. 
In Table \ref{tab:g2_LMDV}, we present the results: similarly to the other models, we find that somewhat large cut-off dependence exists. 
We also find the $h_2$ dependence in the HVP contribution to be of the similar size.  Although comparing the results obtained by using two values of $h_5$ is difficult due to these dependences, 
the results in this work are systematically smaller than those obtained by using the value of $h_5$ in the HLBL calculation 
and are consistent with results in the other models in Table \ref{tab:g2_table}. 
Because the standard method to calculate the HVP contribution to the muon $g-2$ is based on the dispersion relation, 
comparing our results with those obtained by using dispersive approaches is worthwhile.


\begin{table}[h]
\centering
\caption{HVP contributions from the anomalous transition, $a_{\mu}^{\textrm{HVP}:\pi^{0}\gamma}$,
in unit of $10^{11}$ 
for the WZW, VMD, LMD and AdS/QCD models.\\}
\begin{tabular}{c|c|c|c}
\hline
Model&$\Lambda_{\textrm{cut}}=1~{\rm GeV}$&$\Lambda_{\textrm{cut}}=2~{\rm GeV}$&$\Lambda_{\textrm{cut}}=3~{\rm GeV}$\\ \hline
WZW&1.61&2.39&2.82 \\ \hline
VMD&2.54&3.32&3.53 \\ \hline
LMD&2.71&4.51&6.02 \\ \hline
AdS/QCD&2.27&2.76&2.78 \\ \hline
\end{tabular}
\label{tab:g2_table}
\end{table}

\begin{table}[th!]
\centering
\caption{HVP contributions from the anomalous transition, $a_{\mu}^{\textrm{HVP}:\pi^{0}\gamma}$, in units of $10^{11}$  
for the LMD+V model.\\}
\begin{tabular}{c||c|c|c||c|c|c}
\hline 
$h_2$ ($\text{GeV}^2$)&\multicolumn{3}{|c||}{$h_5=6.93\, \text{GeV}^4$
                        \cite{Knecht:2002}}&\multicolumn{3}{|c}{$h_5=3.38\,
                                               \text{GeV}^4$ [This work]} \\ \hline
&$\Lambda_{\textrm{cut}}=1~{\rm GeV}$&$\Lambda_{\textrm{cut}}=2~{\rm GeV}$&$\Lambda_{\textrm{cut}}=3~{\rm GeV}$&
$\Lambda_{\textrm{cut}}=1~{\rm GeV}$&$\Lambda_{\textrm{cut}}=2~{\rm GeV}$&$\Lambda_{\textrm{cut}}=3~{\rm GeV}$ \\ \hline
$10$&4.15&7.68&8.87&3.25&4.77&5.15 \\ \hline
$0 $&3.38&5.77&6.48&2.69&3.53&3.66 \\ \hline
$-10.63 $&2.55&3.66&3.82&2.07&2.13&1.95 \\ \hline
\end{tabular}
\label{tab:g2_LMDV}
\end{table}

\begin{figure}[h]
\centering
\includegraphics[width=0.6\textwidth]{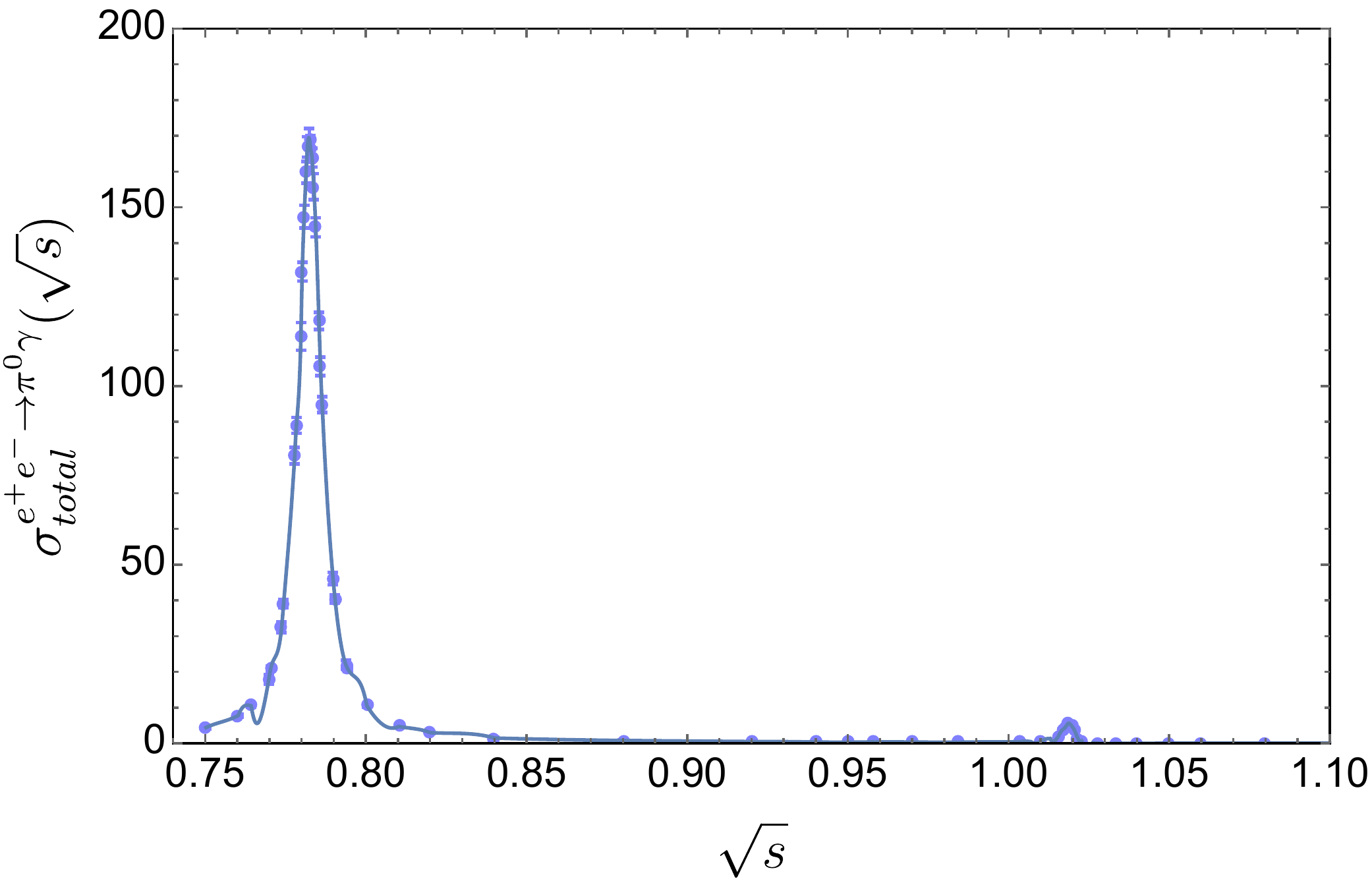}
\caption{
The solid line denotes the interpolated cross-section to fit the SND data, which are shown as dots. 
}
\label{data}
\end{figure}

One can estimate the HVP contributions due to the anomalous transition $\gamma^*\rightarrow \pi^0\gamma^*$ from the dispersion relation~\cite{Krokhin}, 
\begin{equation}
a_{\mu}^{\rm HVP:\pi^{0}\gamma}=\frac{1}{4\pi^3}\int_{s_{\rm th}}^{\infty}{\rm d}s\,\sigma_{\textrm{total}}^{e^+ e^- \rightarrow \pi^0 \gamma}\cdot\left(\frac{m_{\mu}^2}{3s}K(s)\right),
\label{dis_relation}
\end{equation}
where we take $s_{\rm th}=4m_{\mu}^2$, since the contributions from $s<4m_{\mu}^2$ are negligible, and \begin{equation}
K(s)=\frac{3s}{m_{\mu}^2}\left[\frac{x^2}{2}(2-x^2)+\frac{(1+x^2)(1+x)^2}{x^2}\left\{\ln(1+x)-x+\frac{x^2}{2}\right\}+\frac{1+x}{1-x}\,x^2\ln x\right]
\label{dis}
\end{equation}
with $x\equiv(1-\beta_{\mu})/(1+\beta_{\mu})$ and $\beta_{\mu}\equiv\sqrt{1-4m_{\mu}^2/s}$.
By using the interpolating function constructed from the SND data (the solid line in Figure~\ref{data}) 
for $\sigma_{\textrm{total}}^{e^+ e^- \rightarrow \pi^0 \gamma}$ in Eq.~\ref{dis_relation}, we obtain 
\begin{equation}
a_{\mu}^{\rm HVP:\pi^{0}\gamma}=4.50\times10^{-10},
\end{equation}
which turns out to be 
an order of magnitude larger than the one obtained from the two-loop
calculation by using the anomalous form factor. Because the differences among the considered models are much smaller than the differences with the dispersive results, 
determining which model is the most appropriate one for calculating the HLBL contributions to the muon $g-2$ is difficult. 
The corrections from both the HLBL and the HVP involving the pion-photon-photon transition form factor are expected to be of the same order, $\alpha^3$. 
Such a naive expectation seems to be correct if we consider the dispersive results for the HVP and the HLBL contributions obtained in the pion-pole approximation. 
However, our results for the HVP contribution obtained using the transition form factor turn out to be systematically smaller than those results obtained using the dispersion relation in Eq.~\ref{dis_relation} by a factor of ten. 
We leave this discrepancy to a future study. 
In addition, computing the HLBL corrections by using the parameters $h_5$, $M_{V_1}$ and $M_{V_2}$  found in our study would be interesting because, 
as shown in Table \ref{tab:g2_LMDV}, the HVP contributions to muon $g-2$ for different values of $h_5$ are quite different.

\begin{center}
\begin{acknowledgments}
This work was supported by a two-year research grant from Pusan National University.
\end{acknowledgments}
\end{center}



\end{document}